\documentclass[twocolumn,showpacs,preprintnumbers,amsmath,amssymb,prb,aps]{revtex4-1}
\usepackage{epsfig}
\usepackage{epstopdf}
\usepackage{graphicx}
\usepackage{dcolumn}
\usepackage{bm}
\usepackage{textcomp}
\usepackage{array}
\usepackage{subcaption}
\usepackage{booktabs}

\begin{document}

\title{Understanding the Seebeck coefficient of LaNiO$_{3}$ compound in the temperature range $300-620$ K}
\author{Arzena Khatun$^{1,}$}
\altaffiliation{Electronic mail: khatunarzena@gmail.com}
\author{Shamim Sk$^{1}$}
\author{Sudhir K. Pandey$^{2,}$}
\altaffiliation{Electronic mail: sudhir@iitmandi.ac.in}
\affiliation{$^{1}$School of Basic Sciences, Indian Institute of Technology Mandi, Kamand - 175075, India}
\affiliation{$^{2}$School of Engineering, Indian Institute of Technology Mandi, Kamand - 175075, India}
\date{\today} 

\begin{abstract}

Transition metal oxides have been attracted much attention in thermoelectric community from the last few decades. In the present work, we have synthesized LaNiO$_{3}$ by a simple solution combustion process. To analyze the crystal structure and structural parameters we have used Rietveld refinement method wherein FullProf software is employed. The room temperature x-ray diffraction indicates the trigonal structure with space group $R \, \overline{3} \, c$ (No. 167). The refined values of lattice parameters are a = b = 5.4615 \AA\, $\&$ c = 13.1777 \AA. Temperature dependent Seebeck coefficient (S) of this compound has been investigated by using experimental and computational tools. The measurement of S is conducted in the temperature range $300-620$ K. The measured values of S in the entire temperature range have negative sign that indicates \textit{n}-type character of the compound. The value of S is found to be $\sim$ $-$8 $\mu$V/K at 300 K and at 620 K this value is $\sim$ $-$12 $\mu$V/K. The electronic structure calculation is carried out using DFT+\textit{U} method due to having strong correlation in LaNiO$_{3}$. The calculation predicts the metallic ground state of the compound. Temperature dependent S is calculated using BoltzTraP package and compared with experiment. The best matching between experimental and calculated values of S is observed when self-interaction correction is employed as double counting correction in spin-polarized DFT + \textit{U} (= 1 eV) calculation. Based on the computational results maximum power factors are also calculated for \textit{p}-type and \textit{n}-type doping of this compound. 

\vspace{0.2cm}
\end{abstract}
 
\maketitle

\section{Introduction}
 
The demand of electricity has highly been increased day by day to fulfill the daily base requirement of mankind. Due to having poor conversion efficiency of engines most of the consumed energy is directly released to the environment in the form of waste heat energy\cite{waste1, waste2} that enhances the temperature of the atmosphere. So, there is a need of searching new energy sources to accomplish the demand of mankind's electricity without hampering the ecological and environmental conditions. One conversion system named thermoelectric generator (TEG) has the potential capability to transform the waste heat energy into electrical energy\cite{electrical1, electrical2} and that will help in controlling the global warming also. A dimensionless parameter figure-of-merit (ZT)\cite{ZT} defines the efficiency of TEG, which is expressed as 
\begin{equation}
ZT = \frac{S^{2}\sigma T}{\kappa}
\end{equation}
where, the symbols S, $\sigma$, T and $\kappa$ denote the Seebeck coefficient, electrical conductivity, absolute temperature and thermal conductivity of the material, respectively. For efficient TE material it is required to maximize it's electrical power factor ($S^{2}\sigma$), PF as well as to minimize the $\kappa$. But, it is difficult to optimize the electrical PF and $\kappa$ at the same time because there exists a strong correlation among S, $\sigma$ and $\kappa$\cite{chen, shamim_mrx}. Many experimental and computational approaches are used to overcome this difficulty. 

So far, bismuth and lead telluride alloys are the best example of commercially available TE materials\cite{led1, led2} mostly used for making refrigerator and thermoelectric generator. In spite of having high demand, these materials are not suitable for high temperature applications because they are unstable at elevated temperature. Further, having environmentally unfavourable issues such as oxidized in the air, toxic in nature and high fabrication cost of device make limitation in their use. Metal oxides create a new hope for the researchers to overcome these problems and are believed good candidates of high temperature TE applications\cite{app1, app2, app3}. Metal oxides such as ZnO\cite{zno}, Ca-Co-O\cite{cao} and Na$ _{x} $CoO$ _{2} $\cite{nao} are the example of incentive TE materials in oxide family.

Studying the properties of strongly correlated electron systems (SCES) have been attracted much attention of scientific community from last few decades as they show unusual properties such as metal-insulator (MI) transition, half-metallicity, S, $\sigma$, high temperature superconductivity, colossal magnetoresistance etc\cite{half1, half2, half3, half4}. In the search of TE materials for high temperature application, SCES were found more suitable candidates in TE field \cite{terasaki,maignan,saurabh_lco,shamim_nco}. Correlation effects in SCES were shown to enhance the S \cite{terasaki,maignan,saurabh_lco,oudovenko,boehnke,mravlje}. The spin and orbital degrees of freedom have been found to play an important role for large values of S in this system\cite{koshibae1,koshibae2,maekawa}. The experimental and theoretical studies of S on SCES have been carried out by many groups\cite{terasaki,maignan,saurabh_lco,shamim_nco,bocelli,sales,tomczak,senaris,jirk,kawata}. LaNiO$_{3}$ is a strongly correlated transition metal oxide, which comes under rare earth nickelates (RNiO$_{3}$, R = rare earth) group. The existence of correlation in LaNiO$_{3}$ is indicated by the magnetic susceptibility, resistivity at low temperatures and heat capacity data reported by Sreedhar \textit{et al}\cite{sreedhar}. Many research groups have worked on rare earth nickelates and show that all compounds of this group have MI transition at lower temperatures\cite{insul1, insul2} and their ground state is insulating antiferromagnetic except LaNiO$_{3}$. The absence of magnetic order in LaNiO$_{3}$ compound breaks the trend. For LaNiO$_{3}$ compound researchers reported that it has metallic and paramagnetic ground state\cite{insul1, insul2, para1, Zhou}. Recently, one group has studied the properties of strongly correlated LaNiO$_{3}$ compound and revealed that it has high metallicity and antiferromagnetic correlations\cite{anti}. LaNiO$_{3}$ has controversial magnetic and transport properties.

Although the TE properties of LaNiO$_{3}$ have been studied in the wide temperature range\cite{Gayathri,Hsiao,Zhou,tak}. But, a systematic understanding of S in LaNiO$_{3}$ using experimental and theoretical tools is still lacking from the literatures. This gives a motivation to understand the high temperature S of LaNiO$_{3}$ using combined experimental and theoretical tools. We have prepared LaNiO$_{3}$ sample by a simple solution combustion process and investigate the S in the temperature region $300-620$ K. Then experimental value of S is understood with the help of theoretical approach. Theoretical tools always help to give the proper explanation of the experimental results to understand the physical properties of the sample. Simple density functional theory (DFT) \cite{kohn1,kohn2} is not enough to give the electronic structure and physical properties of the SCES. For such a system, DFT + \textit{U} was found as an useful tool to describe the electronic structure and dependent physical properties. In particular, DFT + \textit{U} method has been used to study the electronic structure and S of cobaltate systems  as reported previously\cite{saurabh_lco,saurabh_lbco1,saurabh_lbco2}. Therefore, having correlation in LaNiO$_{3}$ and similar to cobaltate systems, we have used DFT + \textit{U} to understand the experimentally observed S. Based on the theoretical result maximum PF of this compound are also calculated corresponding to \textit{p}-type and \textit{n}-type doping of this compound.     
    
In this work, we have studied the temperature dependent S of LaNiO$_{3}$ compound in the temperature range $300-620$ K by using experimental and theoretical tools. The LaNiO$_{3}$ compound has been synthesized by a simple solution combustion process. The room temperature x-ray diffraction confirms the trigonal structure with space group $R \, \overline{3} \, c$ (No. 167). The refined values of lattice parameters of the unit cell are a = b = 5.4615 \AA\, $\&$ c = 13.1777 \AA. At room temperature the experimental value of S is found to be $\sim$ $-$8 $\mu$V/K and at 620 K the measured value is $\sim$ $-$12 $\mu$V/K. The \textit{n}-type behaviour of the compound in the full studied temperature range is indicated by the negative sign of S. Spin-polarized calculation in DFT+\textit{U} (= 1 eV) method gives the best matching with the experimental result when self-interaction correction is used as a double counting method. We have also calculated the maximum PF for both \textit{n}-type and \textit{p}-type doping and the calculated values are found to be $\sim$ 15.7 $\times$ 10$^{14}$$\mu$WK$^{-2}$cm$^{-1}$s$^{-1}$ and $\sim$ 33.5 $\times$ 10$^{14}$$\mu$WK$^{-2}$cm$^{-1}$s$^{-1}$, respectively at 1200 K.

\section{Experimental and computational details}

LaNiO$_{3}$ sample was synthesized by a simple solution combustion process. The initial ingredient used for synthesis was a solution of chitosan. For making this solution 1 g of chitosan was mixed in the mixture of 50 mL deionized water with 2$\%$ acetic acid. After that, appropriate amount of La(NO$_{3}$)$_{3}$.6H$_{2}$O and Ni(NO$_{3}$)$_{2}$.6H$_{2}$O nitrates were added in 10 mL of as prepared chitosan solution. The mixture was magnetically stirred for 30 minutes so that the nitrates completely dissolved with chitosan solution and form a viscous mixture. The viscous mixture was placed on a hot plate and heated up to $\sim$ $200 ^{\circ}$ C with continue stirring. Consequently, the chitosan and nitrates were completely decomposed and makes the primary powder. This primary powder were then placed in a furnace to heat up to $850 ^{\circ}$ C for 5 hours. The temperature increment rate was set to be 2$ ^{\circ}$ C/min. Finally obtained powder were pressed into circular pellet and kept at $500 ^{\circ}$ C for 24 hours to harden the pellet.
 
x-ray diffraction (XRD) of the sample was done at room temperature in order to examine the crystal structure of the sample. In XRD, Cu K$\alpha$ radiation ($\lambda$ = 1.5418 \AA) was used for producing the diffractogram. The resultant diffractogram was recorded in the 2$\theta$ range, $20^{\circ}-90^{\circ}$ with step counting mode 0.02. Rietveld refinement method with FullProf software was used to analyze the crystal structure and the structural parameters.

Temperature dependence measurement of S is done using home-made instrumental setup\cite{Setup}. The experiment is carried out in the temperature range $300-620$ K. For the measurement circular shape pellet having 8 ($\pm$0.1) mm diameter and 0.62 ($\pm$0.02) mm thickness is used.  

By using DFT \cite{kohn1,kohn2} and DFT+\textit{U} methods, we have calculated the electronic structure to analyze the experimental result of the sample. The calculations are done in WIEN2k code \cite{Wien2k} wherein, full-potential linearized augmented plane wave (FP-LAPW) method is employed. The electronic structure calculations are performed self-consistently. To achieve  self-consistency in calculation the convergence criteria of the total energy is set to be less than 0.1 mRy/cell. An exchange correlation functional named PBEsol\cite{Pbesol} is used for calculation. The muffin-tin sphere radii are fixed to 2.50, 1.95 and 1.68 Bohr for La, Ni and O atoms, respectively. To calculate the ground-state  electronic structure of the sample the experimental lattice parameters a = b = 5.4615 \AA\, $\&$ c = 13.1777 \AA\, of trigonal structure corresponds to $R \, \overline{3} \, c$ (No. 167) space group are taken as initial input values. Self-interaction correction (SIC)\cite{SIC1,SIC2} and around mean field (AMF)\cite{AMF} methods are used as double counting correction in DFT+\textit{U}. The on-site coulomb interaction srength, \textit{U} is taken as 1 and 2 eV for Ni 3\textit{d} electrons in order to explain the experimental result properly. BoltzTraP package\cite{boltztrap} is used to calculate the temperature dependence S of the sample. The BoltzTraP package\cite{boltztrap} is based on constant relaxation time approximation and interfaced with WIEN2k\cite{Wien2k}. 40 $\times$ 40 $\times$ 40 k-points are used in the full Brillouin zone to calculate the temperature dependence S more accurately.       
       
\section{Results and Discussion}

Fig. 1. presents the room-temperature x-ray diffractogram of LaNiO$_{3}$. Rietveld refinement has been employed to analyze the XRD pattern. Pseudo-Voigt function is used in the refinement process to fit the shapes of the Bragg peaks. The ratio of weighted profile factor (R$_{wp}$) and expected weighted profile factor (R$_{exp}$) is the indicator of goodness of fit of crystal structure and for our sample this value is 1.71. The result of refinement confirms that LaNiO$_{3}$ sample has trigonal structure, with space group $R \, \overline{3} \, c$ (No. 167). The refined values of lattice parameters of the unit cell are a = b = 5.4615 \AA\, $\&$ c = 13.1777 \AA\,, which are in well agreement with the reported values\cite{munoz}. 

Debye Scherrer formula is used to calculate the crystallite size (D) of the sample. The expression of the Debye Scherrer formula is 
\begin{equation}
D = \frac{0.94\lambda}{Bcos\theta}
\end{equation}
wherein, $\lambda$ is the  wavelength of Cu K$\alpha$ radiation of x-ray source (1.5418 \AA), \textit{B} is the line broadening at full width half maxima, in radians and $\theta$ is the Bragg's angle. By using the above expression we have calculated the crystallite size of LaNiO$_{3}$. The calculated value of crystallite size is found to be $\sim$ 17.6 nm in the present case.

Fig. 2. shows the crystal structure of LaNiO$_{3}$ compound. The conventional unit cell of this compound has 30 atoms wherein, the Wyckoff positions of La, Ni and O atoms are 6a(0, 0, 0.25), 6b(0, 0, 0) and 18e(0.5468, 0, 0.25), respectively. In a conventional unit cell there exists three primitive unit cells.

\begin{figure} 
\includegraphics[width=0.85\linewidth, height=6.8cm]{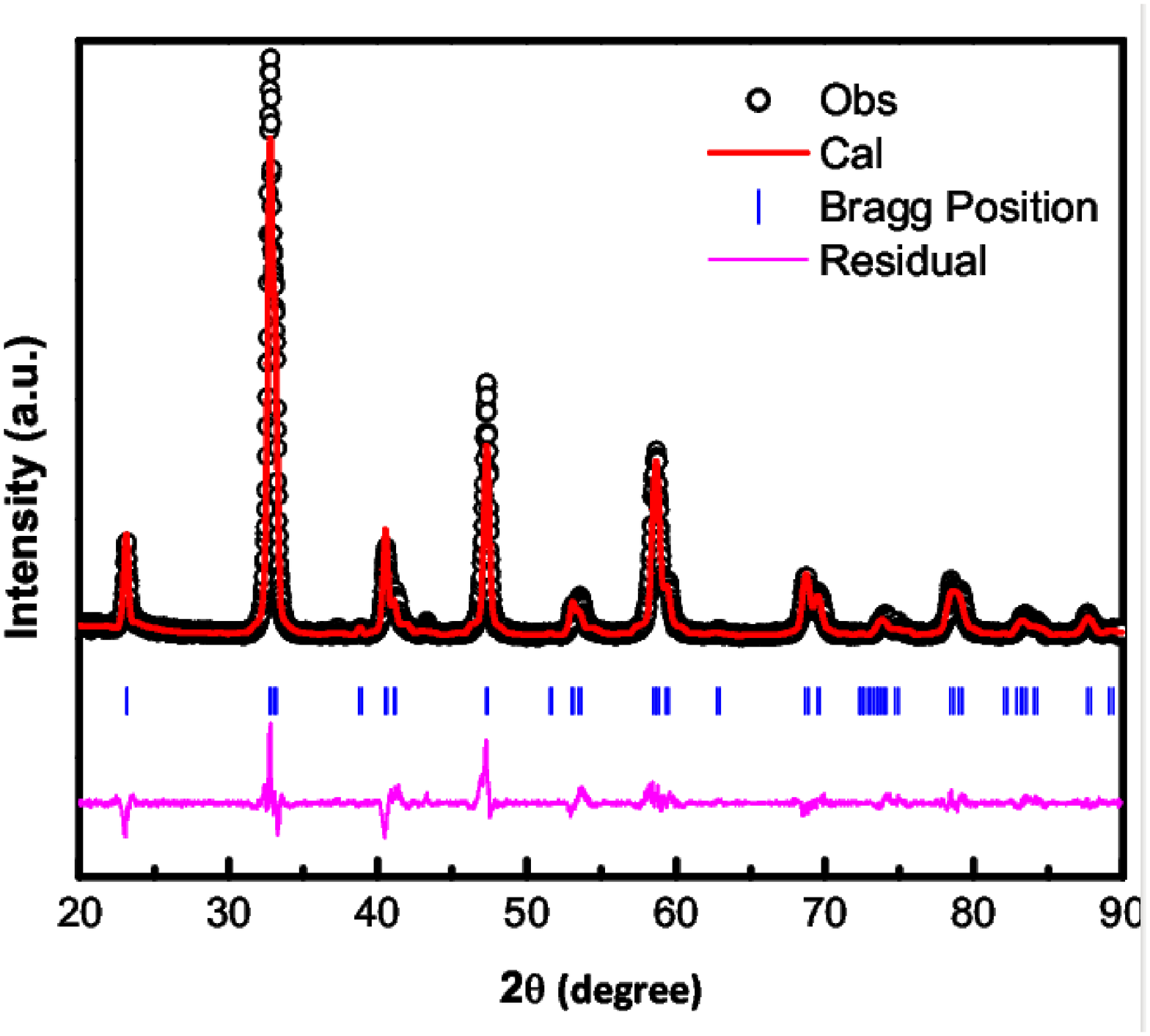} 
\caption{\small{Room temperature x-ray diffraction of LaNiO$_{3}$.}}
\end{figure} 
    
\begin{figure} 
\includegraphics[width=0.75\linewidth, height=6.5cm]{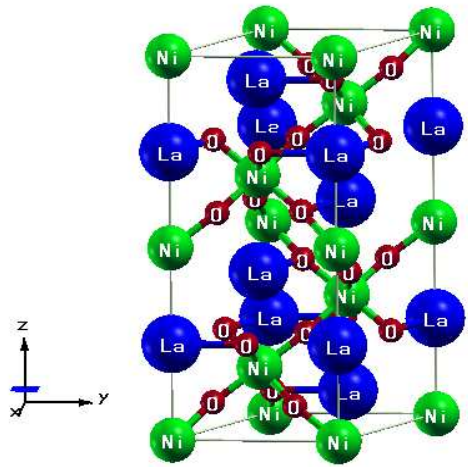} 
\caption{\small{Crystal structure of LaNiO$_{3}$.}}
\end{figure} 

\begin{figure} 
\includegraphics[width=0.90\linewidth, height=6.5cm]{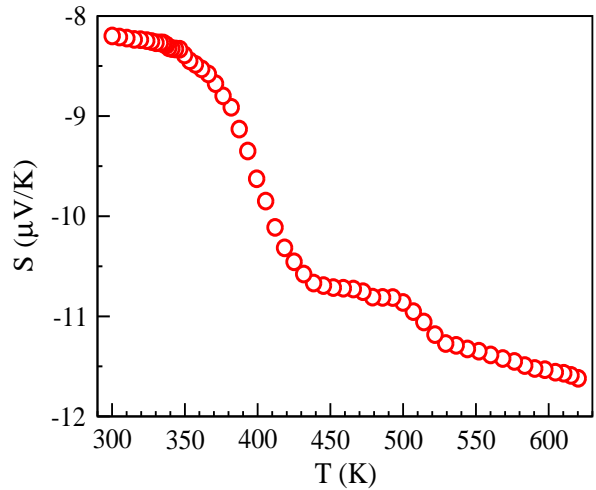} 
\caption{\small{Temperature dependent Seebeck coefficient (S) of LaNiO$_{3}$ compound.}}
\end{figure} 

\begin{figure*}
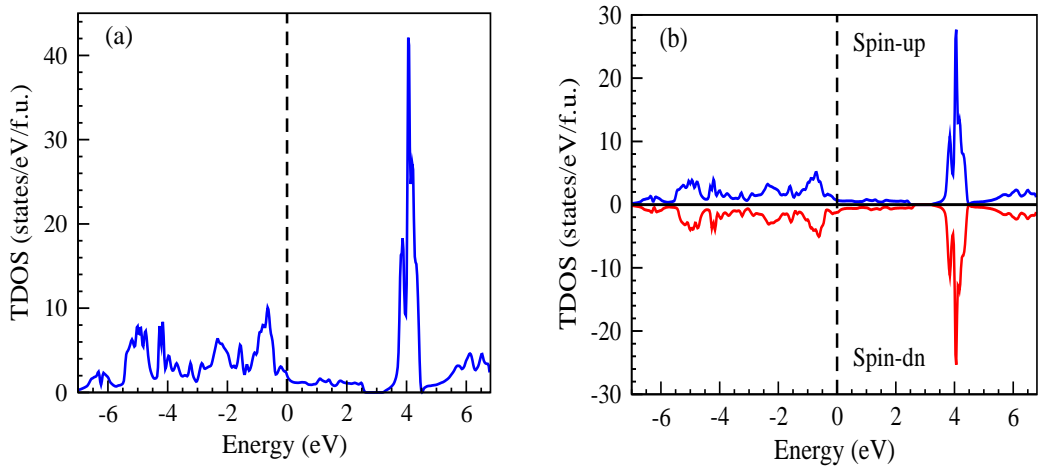
 
\begin{subfigure}{0.40\textwidth}
\includegraphics[width=0.90\linewidth, height=6.0cm]{fig4_1.eps} 
\end{subfigure}
\begin{subfigure}{0.40\textwidth}
\includegraphics[width=0.90\linewidth, height=6.2cm]{fig4_2.eps} 
\end{subfigure}
\caption{\small{Total density of states (TDOS) of (a) spin-unpolarized (SUP) calculation and (b) spin-polarized (SP) calculation using DFT.}}
\end{figure*}

\subsection{\label{sec:level2}Experimental transport properties}  
       
The result of experimentally observed temperature dependent S of LaNiO$_{3}$ compound is shown in Fig. 3. The S measurement is conducted in the temperature range $300-620$ K. The room temperature value of S is found to be $\sim$ $-$8 $\mu$V/K. Gayathri \textit{et al.}\cite{Gayathri} reported the S of well oxygenated LaNiO$_{3}$ film in the range of $-$15 to $-$20 $\mu$V/K at 300 K. Hsiao \textit{et al.}\cite{Hsiao} also studied the S of nanostructured LaNiO$_{3-x}$ ($x$ $>$ 0) and the obtained room temperature value was in the range of $-$12 to $-$15 $\mu$V/K. Therefore, it is seen that the reported room temperature values of S \cite{Gayathri, Hsiao} are lower in comparison to measured value of our sample. The synthesis conditions of our sample is different from that of Gayathri \textit{et al.}\cite{Gayathri} and Hsiao \textit{et al.}\cite{Hsiao}. Usually, the quality of transition metal oxide strongly depends on the synthesis conditions such as pressure \cite{Gayathri, Zhou}, strain, variation in temperature \cite{Hsiao} etc. For which reason the physical properties of oxide thermoelectric materials vary. Therefore, one possible reason for this observed difference in S may be the different synthesis conditions of our sample to the reported works.  Moreover, the variation in S occurs in the content of oxygen if the oxide samples are prepared in different conditions \cite{Gayathri, Hsiao}. May be the different oxygen off-stoichiometry is also responsible for the different values in S. Fig. 3 shows that after 300 K the value of S decreases with temperature up to the highest studied temperature. At 620 K the measured value of S is $\sim$ $-$12 $\mu$V/K. From figure it is clear that the values of S in the entire temperature range are negative that indicates the \textit{n}-type behaviour of the compound. Therefore, the major contribution in S for LaNiO$_{3}$ compound comes from electrons in the full temperature region. 
   
\subsection{\label{sec:level2} Electronic structure calculations}
 
Electronic structure calculation helps to understand the temperature dependent transport properties of the compound. To find out the ground state electronic structure of LaNiO$_{3}$ the spin-unpolarized (SUP) and spin-polarized (SP) calculations were done on this compound within density functional theory (DFT). For SP the calculated value of total converged energy  is found to be $\sim$ 0.27 meV/f.u. lower than that of SUP. This result indicates that the ground state of this compound is magnetic. The magnetic moment per formula unit is calculated as $\sim$ 0.92 $\mu_{B}$. Almost 90$\%$ magnetic moment of the total value comes from Ni atom. So, the dominant contributing atom in magnetic moment is Ni. 
 
Total density of states (TDOS) calculation corresponds to SUP and SP was carried out using DFT and DFT+\textit{U} to explain the experimentally observed S of the LaNiO$_{3}$ compound. The calculated TDOS plots in DFT for both SUP and SP are shown in Fig. 4. In plot, 0 eV represents the Fermi level (E$_{F}$) of the system, which is denoted by the vertical dashed line. Fig. 4 (a) exhibits the TDOS plot of SUP calculation and the obtained value of DOS is  $\sim$ 1.99 states/eV/f.u. at E$_{F}$. According to Stoner theory, this considerable amount of DOS at Fermi level gives an indication of magnetic ground state. Fig. 4 (b) represents the TDOS calculated for SP. From fig. 4 (b) it is clear that, both spin-up and spin-dn channels have finite DOS at Fermi level. For spin-up  channel the value of DOS is found to be $\sim$ 0.67 states/eV/f.u. whereas, for spin-dn channel the value is $\sim$ 1.22 states/eV/f.u. at E$_{F}$. Therefore, these finite but different amount of DOS corresponds to spin-up and spin-dn channels predict the metallic magnetic ground state of the compound. This result is accordance with the reported data\cite{anti}. 

The experimentally observed results of strongly correlated electron systems (SCES) can be described in a better way by using DFT+\textit{U}. LaNiO$_{3}$ is a SCES for that we have used DFT+\textit{U} method to calculate the TDOS. In DFT+\textit{U} method the on-site coulomb interaction strength, \textit{U} was taken 1 and 2 eV for Ni 3\textit{d} electrons to explain the experimental transport property of the sample. To avoid double counting interaction in calculation some methods are embedded in DFT+\textit{U}. But, which method should be used as double counting correction in DFT + \textit{U} to study the TE properties of the materials is missing from the literature. Using any one of the method may not give the accurate TE properties of the materials. Thus, a systematic study in this direction is necessary. Here, we have explored two well known methods naming self-interaction correction (SIC)\cite{SIC1,SIC2} and around mean field (AMF)\cite{AMF} to calculate the TDOS of LaNiO$_{3}$. Fig. 5 presents the calculated TDOS plot of LaNiO$_{3}$ compound corresponding to SP using DFT+\textit{U} method. It is evident from Fig. 5 that both channels of  SIC $\&$ AMF corresponds to \textit{U} = 1 eV give finite DOS at Fermi level that indicates metallic behaviour of the system. On the other hand AMF corresponds to \textit{U} = 2 eV shows that both channels have finite DOS at E$_{F}$ but exception occurs at spin-dn channel for SIC corresponds to \textit{U} = 2 eV. The spin-dn channel corresponds to \textit{U} = 2 eV SIC gives zero DOS at Fermi level that predicts semiconducting character. So, the obtained result of \textit{U} = 2 eV SIC method indicates that the compound has half-metallic characteristic. However, due to having different feature of DOS around $E_{F}$ corresponding to DFT and DFT + \textit{U}, one can expect that these two methods will give the different transport properties of this compound. Moreover, both the double counting corrections in DFT + \textit{U} also give the different features of DOS around $E_{F}$ which is expecting to lead the different results in transport properties.     

\begin{figure} 
\includegraphics[width=0.78\linewidth, height=6.0cm]{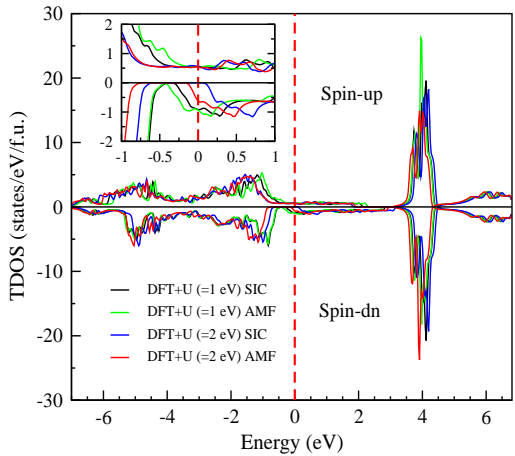} 
\caption{\small{Total density of states (TDOS) using spin-polarized (SP) DFT+\textit{U} (\textit{U} = 1, 2 eV) calculation. Here, SIC and AMF stand for self-interaction correction and around mean field, respectively, which are double counting correction methods  in DFT + \textit{U}.}}
\end{figure}

\begin{figure} 
\includegraphics[width=0.80\linewidth, height=6.0cm]{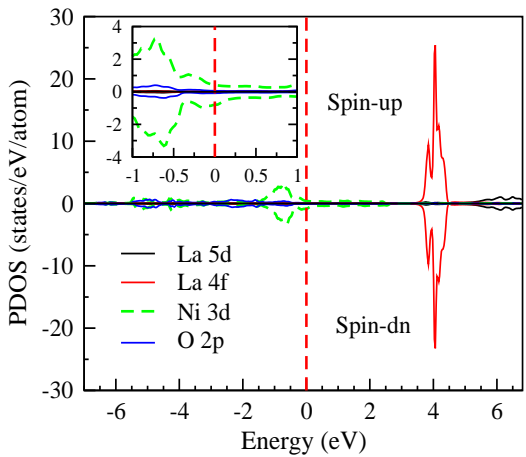} 
\caption{\small{Partial density of states (PDOS) of spin-polarized (SP) DFT calculation.}}
\end{figure}

To see the contribution in TDOS from different atoms of LaNiO$_{3}$, we have carried out the partial density of states (PDOS) calculation. Different atomic states around $E_{F}$ participate in transport properties. We have calculated the PDOS for the constituent atoms (La, Ni and O) of LaNiO$_{3}$. Fig. 6 represents the PDOS corresponds to SP within DFT. The PDOS plot shows that in occupied band (OB) a peak originates near the Fermi level ($\sim$ -2 eV to 0 eV) corresponds to Ni 3\textit{d} orbital. In unoccupied band (UB) a peak with high intensity exists in both the channels and locates in the high energy region. Around $E_{F}$ for both the OB and UB the major contribution in DOS comes from Ni 3\textit{d} orbital with negligibly small contribution from O and La for both the channels (See inset of Fig. 6). The DOS at $E_{F}$ for Ni 3\textit{d} orbital has the value of $\sim$ 0.41 states/eV/atom for spin-up channel whereas, this value is $\sim$ 0.79 states/eV/atom for spin-dn channel. These finite amount of DOS at $E_{F}$ for both the channels are responsible for lowering the S value of LaNiO$_{3}$.

\begin{table*}
\caption{\label{tab:table1}%
\small{Experimental and calculated values of S at different temperature by using DFT and DFT+$U$ methods. The average of absolute deviation of calculated S from experiment is also include in the table. (Here, S is in $\mu$V/K unit). 
}}
\begin{ruledtabular}
\begin{tabular}{lcccccc}
\textrm{Methods}&
\textrm{S at 300 K}&
\textrm{S at 400 K}&
\textrm{S at 500 K}&
\textrm{S at 600 K}&
\textrm{S at 620 K}&
\textrm{$(\sum|S_{exp}-S_{cal}|)/N$}\\
      
\colrule
Exeriment & -8.20 & -9.65 & -10.86 & -11.54 & -11.62 & --- \\
DFT(= 0 eV SIC)      & -14.86 & -19.47 & -23.99 & -28.36 & -29.20 & 11.92 \\
DFT+$U$(= 1 eV SIC)    & -6.59 & -8.64 & -10.65 & -12.64 & -13.03 & 0.84 \\
DFT+$U$(= 1 eV AMF)    & -7.11 & -9.59 & -12.12 & -14.62 & -15.11 & 1.17 \\
DFT+$U$(= 2 eV SIC)    & 1.71 & 2.13 & 1.97 & 1.29 & 1.11 & 12.05 \\
DFT+$U$(= 2 eV AMF)    & -6.58 & -8.35 & -9.75 & -10.92 & -11.14 & 1.13 \\
\end{tabular}
\end{ruledtabular}
\end{table*}

\begin{figure} 
\includegraphics[width=0.80\linewidth, height=6.0cm]{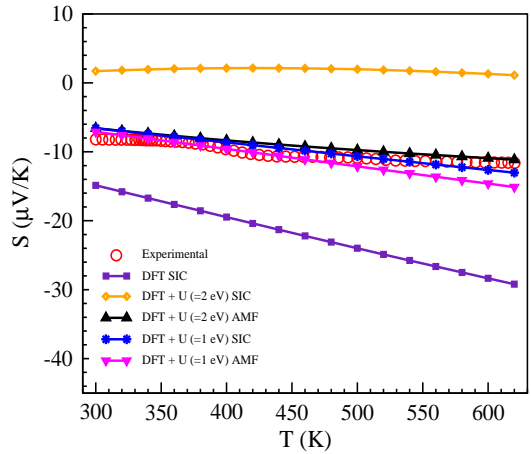} 
\caption{\small{Experimental and calculated (using DFT+\textit{U}) values of Seebeck coefficient (S) as a function of temperature. Self-interaction correction (SIC) and around mean field (AMF) are double counting correction methods in DFT + \textit{U}.}}
\end{figure}

\begin{figure*} 
\includegraphics[width=.80\linewidth, height=6.5cm]{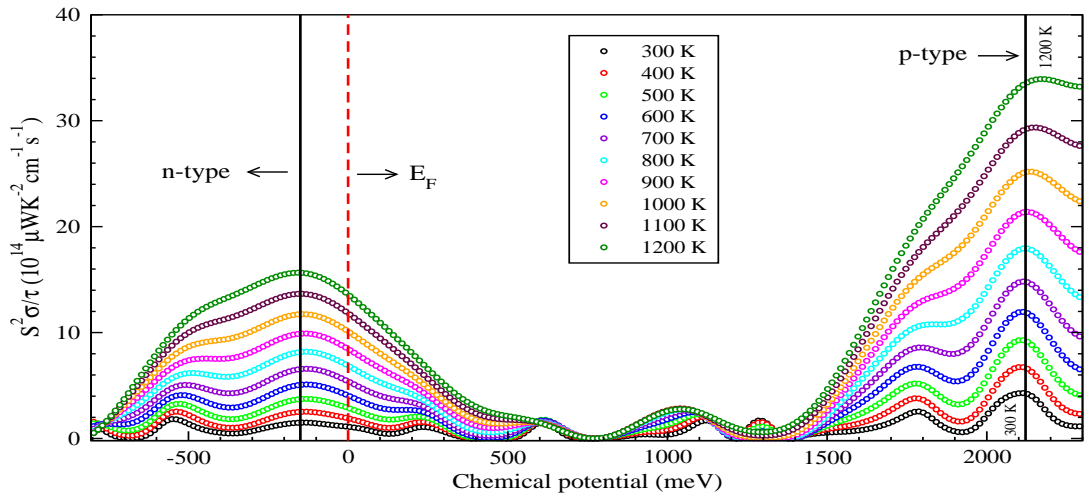} 
\caption{\small{Variation of power factor per relaxation time as a function of chemical potential at different temperatures.}}
\end{figure*}

\subsection{\label{sec:level2}Calculated transport properties}
 
In order to examine the experimentally observed S data of LaNiO$_{3}$ compound it is required to calculate the total S value from spin-up and spin-dn channels. Using BoltzTraP package\cite{boltztrap} we have calculated the temperature dependent S of LaNiO$_{3}$ compound corresponding to spin-up and spin-dn channels. Based on two current model the total temperature dependence S was calculated. Two current model gives an expression for calculating S as \cite{current1, current2, current3, current4}
\begin{equation}
S = \frac{(\sigma^{\uparrow}S^{\uparrow}+\sigma^{\downarrow}S^{\downarrow})}{(\sigma^{\uparrow}+\sigma^{\downarrow})}
\end{equation}
In the above equation, $\sigma^{\uparrow}$ and $\sigma^{\downarrow}$ denote the electrical conductivities whereas, S$^{\uparrow}$ and S$^{\downarrow}$ denote the Seebeck coefficients of up and dn-channels, respectively. Fig. 7 represents the calculated values of S for SP DFT and DFT+\textit{U} methods. The calculated values are compared with the experiment in the same figure. For DFT, the calculated values of S are $\sim$ $-$15 and $\sim$ $-$29 $\mu$V/K at 300 K and 620 K, respectively. These values of S are far away from the experimental result of the sample as shown in the figure. We have already discussed that simple DFT tool is not suitable to explain the experimental S for such a SCES. So, for the better understanding of experimental S we have performed the calculation using DFT+\textit{U}. \textit{U} are chosen as 1 and 2 eV for Ni 3\textit{d} electrons. In DFT+\textit{U}, we have used SIC and AMF as double counting correction methods to calculate S. At 300 K, the calculated value of S for \textit{U} = 1 eV SIC (AMF) is $\sim$ $-$6.59 ($-$7.11) $\mu$V/K, whereas for \textit{U} = 2 eV SIC (AMF) this value is $\sim$ 1.71 ($-$6.58) $\mu$V/K. At 620 K, the S value for \textit{U} = 1 eV SIC (AMF) is $\sim$ $-$13.03 ($-$15.11), whereas for \textit{U} = 2 eV SIC (AMF) this value is $\sim$ 1.11 ($-$11.14) $\mu$V/K. These values are quite closer to the experimental value that is clearly seen in plot except \textit{U} = 2 eV SIC value. In case of \textit{U} = 2 eV SIC the calculated values of S in whole temperature range are positive and which is inconsistent with our experimental results. From TDOS, it was seen that \textit{U} = 2 eV SIC method predicts the half-metallic behaviour in this compound which was inconsistent with the reported one\cite{anti}. Therefore, different S values are expected in the case of \textit{U} = 2 eV SIC as compared to others.         

\begin{figure} 
\includegraphics[width=0.8\linewidth, height=6.0cm]{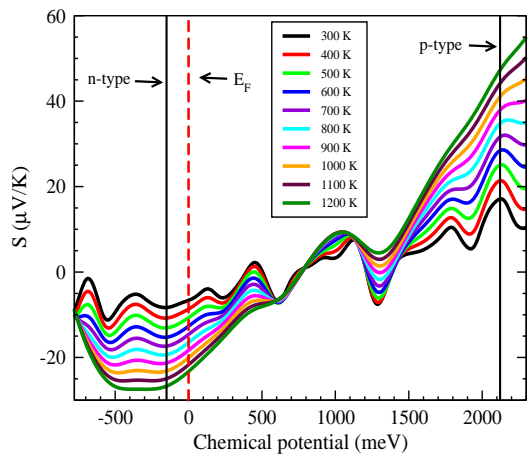} 
\caption{\small{ Variation of Seebeck coefficient (S) with chemical potential at different temperatures.}}
\end{figure}

Table 1 shows the experimental and calculated values of S in the temperature region $300-620$ K. We have calculated the average of absolute deviation of calculated S from experiment in order to find out the best matching between experimental and calculated data. \textit{N} is used as the total number of data points at different temperatures to calculate the average of absolute deviation. From table it is evident that the calculated S values using \textit{U} = 1 eV SIC have the lowest deviation. Fig. 7 also shows that the calculated S values corresponding to \textit{U} = 1 eV SIC gives good match with experimental result in the entire temperature range as compared to others. Therefore, \textit{U} = 1 eV SIC is more appropriate to understand the TE properties of the materials.   

\subsection{\label{sec:level2}Power factor}

In equation (1) the term $S^{2}\sigma$ denotes the power factor (PF) that implies if a material has larger PF then the material will have higher ZT. Based on  PF one can select which material can be used as a good TE material for applications. Here, we have calculated the PF per relaxation time ($\tau$) as a function of $\mu$ at different temperatures (by taking \textit{U} = 1 eV SIC result) as shown in Fig. 8. In figure, vertically dashed line represents the Fermi level, E$_{F}$ of the system that is taken as 0 eV. Here, it should be mentioned that to compare the experimental S we have calculated S at $E_{F}$. Using Fig. 8, we have tried to find out the maximum PF corresponding to \textit{p}-type and \textit{n}-type behaviour of this compound. This prediction then helps to make TEG as for making TEG we need both the \textit{p}-type and \textit{n}-type materials. Keeping these in mind, we found that the maximum PF can be obtained at $\mu$ $\approx$ $-$150 meV (solid vertical line) by electron doping of $\sim$3.6 $\times$ $10^{21}$ $cm^{-3}$. The \textit{n}-type behaviour in this region is clearly seen by the negative sign of S as shown in Fig. 9. The values of PF at this $\mu$ are $\sim$ 1.5 $\times$ 10$^{14}$$\mu$WK$^{-2}$cm$^{-1}$s$^{-1}$ and $\sim$ 15.7 $\times$ 10$^{14}$$\mu$WK$^{-2}$cm$^{-1}$s$^{-1}$ which, correspond to 300 K and 1200 K, respectively. For the \textit{p}-type character of this compound the maximum PF can be found at $\mu$ $\approx$ 2121 meV (solid vertical line), which corresponds to hole doping of $\sim$4.6 $\times$ $10^{22}$ $cm^{-3}$. The \textit{p}-type behaviour in this region is clearly seen in Fig. 9. The maximum PF at $\sim$2121 meV are calculated as $\sim$ 4.3 $\times$ 10$^{14}$$\mu$WK$^{-2}$cm$^{-1}$s$^{-1}$ and $\sim$ 33.5 $\times$ 10$^{14}$$\mu$WK$^{-2}$cm$^{-1}$s$^{-1}$ at 300 and 1200 K, respectively. 
     
\section{CONCLUSIONS}
In conclusion, the combined experimental and computational tools are used to investigate the temperature dependent Seebeck coefficient (S) of LaNiO$_{3}$ compound. The LaNiO$_{3}$ compound was synthesized by a simple solution combustion process. The room temperature x-ray diffraction confirms the trigonal structure with space group $R \, \overline{3} \, c$ (No. 167). Rietveld refinement gives the lattice parameters of a = b = 5.4615 \AA\, $\&$ c = 13.1777 \AA. The measurement of S is carried out in the temperature range $300-620$ K. The experimental values of S are found to be $\sim$ $-$8 $\mu$V/K and $\sim$ $-$12 $\mu$V/K at 300 and 620 K, respectively. The negative values of S through the studied temperature range indicate the \textit{n}-type character of the compound. To know the electronic structure of strongly correlated LaNiO$_{3}$ compound we have carried out the calculation in DFT+\textit{U}. The calculated result predicts the metallic ground state of the compound. By using BoltzTraP package we have also calculated the temperature dependent S values to analyze the experimental result. Spin-polarized DFT + \textit{U} (= 1 eV) calculation gives the best matching with the experimental result when self-interaction correction method is employed as double counting correction. Finally, we have calculated maximum PF for both \textit{n}-type and \textit{p}-type behaviour and the values are found to be $\sim$ 15.7 $\times$ 10$^{14}$$\mu$WK$^{-2}$cm$^{-1}$s$^{-1}$ and $\sim$ 33.5 $\times$ 10$^{14}$$\mu$WK$^{-2}$cm$^{-1}$s$^{-1}$, respectively at 1200 K.


\end{document}